\documentclass{iaus}
\usepackage{graphicx}

\title[The Gama Panchromatic Survey] %% give here short title %%
{The GAMA Panchromatic Survey}

\author[Simon P. Driver]   %% give here short author list %%
{Simon P. Driver$^{1,2}$}

\affiliation{$^3$ International Centre for Radio Astronomy Research, University of Western Australia, Crawley, WA 6009, Australia, \\
$^2$ Scottish Universities Physics Alliance (SUPA), School of Physics and Astronomy, University of St Andrews, St Andrews, Scotland}

\pubyear{2013}
\volume{295}  %% insert here IAU Symposium No.
\pagerange{1--2}
\date{?? and in revised form ??}
\jname{The intriguing life of massive galaxies}
\editors{Daniel Thomas, Anna Pasquali \& Ignacio Ferreras}
\begin{document}

\maketitle

\begin{abstract}
The Galaxy And Mass Assembly Survey (GAMA) has now been operating for
almost 5 years gathering spectroscopic redshifts for five regions of
sky spanning 300 sq degrees in total to a depth of $r<19.8$ mag. The
survey has amassed over 225,000 redshifts making it the third largest
redshift campaign after the SDSS and BOSS surveys. The survey has two
novel features that set it apart: (1) complete and uniform sampling to
a fixed flux limit ($r<19.8$ mag) regardless of galaxy clustering due
to multiple-visits to each sky region, enabling the construction of
high-fidelity catalogues of groups and pairs, (2) co-ordination with
diverse imaging campaigns which together sample an extremely broad
range along the electro-magnetic spectrum from the UV (GALEX) through
optical (VST KIDs), near-IR (VISTA VIKING), mid-IR (WISE), far-IR
(Herschel-Atlas), 1m (GMRT), and eventually 20cm continuum and
rest-frame 21cm line measurements (ASKAP DINGO). Apart from the ASKAP
campaign all multi-wavelength programmes are either complete or in the
final stages of observations and the UV-far-IR data are expected to be
fully merged by the end of 2013. This article provides a brief flavour
of the coming panchromatic database which will eventually include
measurements or upper-limits across 27 wavebands for 380,000
galaxies. GAMA DR2 is scheduled for the end of January 2013.
\end{abstract}

\firstsection % if your document starts with a section,
              % remove some space above using this command.
\section{Introduction}
\begin{figure}[h]
\includegraphics[height=5.5in,width=4.0in,angle=90.0]{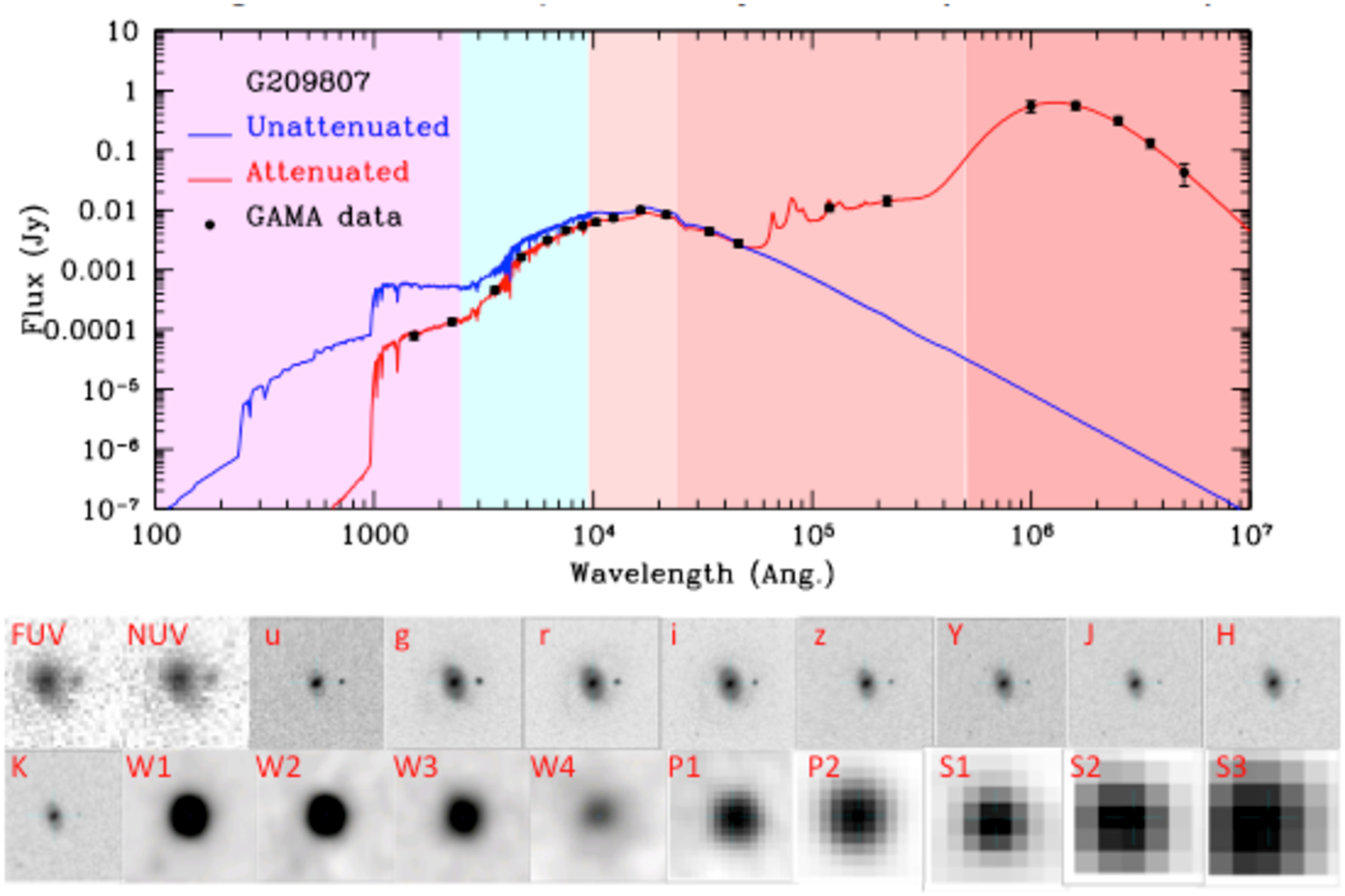}
\caption{({\it upper plot}) MAGPHYS modelling of a single GAMA galaxy
  detected in GALEX, SDSS, UKIDSS, WISE, HERSCHEL PACS and HERSCHEL
  SPIRE. ({\it lower panels}) The postage stamps images are shown in
  the lower panels highlighting the varied spatial resolutions of the
  contributing surveys. \label{fig1}}
\end{figure}

The Galaxy And Mass Assembly Survey (GAMA; \cite{driver09},
\cite{driver11}) will contain spectroscopic redshifts at $>98\%$
completeness for 380,000 galaxies drawn from five distinct sky regions
(2$^h$, 9$^h$, 12$^h$, 15$^h$, and 23$^{h}$). Details of the survey
concept, input catalogue, tiling strategy and first data release are
provided in \cite[Driver et al (2009)]{driver09}, \cite[Baldry et al
  (2010)]{baldry10}, \cite[Robotham et al (2010)]{robotham10}, and
\cite[Driver et al (2011)]{driver11} respectively.  The construction
of the first group catalogue is given by \cite[Robotham et al
  (2011)]{robotham11} and the measurement of robust stellar masses and
S\'ersic profiles by \cite[Taylor et al (2011)]{taylor11} and
\cite[Kelvin et al (2012)]{kelvin12} respectively.  Current early
science includes the measurement of the nearby galaxy luminosity
functions from $UV-K$ \cite[Driver et al (2012a)]{driver12a} and their
evolution in $ugriz$ to $z<0.5$ by \cite[Loveday et al
  (2012)]{loveday12}. The nearby galaxy stellar mass function, which
exhibits a distinctive upturn at intermediate masses, is given by
\cite[Baldry et al (2012)]{baldry12} and upcoming papers by
\cite[Taylor et al (2013)]{taylor13}, \cite[Lopez-Cruz et al
  (2013)]{lopez-cruz13}, \cite[Gunawardhana et al (2013)]{guna13} and
\cite[Robotham et al (2013)]{robotham13} explore galaxy colour
bimodality, the evolution of the mass-metallicity relation, the
evolution of the H$\alpha$ luminosity function, and the life and times
of L$^*$ galaxies. Numerous ($>100$) other papers are currently in
progress. Fig.~\ref{fig1} shows the end-of-year 2012 update to the
survey of surveys table given in \cite[Driver (2012b)]{driver12b} and
provides some indication of GAMAs emerging status amongst established
leading surveys from the past three decades. Modern surveys like GAMA,
BOSS and VIPERS are all expected to rise rapidly up this league table
over the coming years and updates will be reported annually.

\begin{table}[h]
  \begin{center}
  \caption{Major extragalactic optical/near-IR surveys and their impact in terms of refereed papers and citations}
  \label{tab1}
 {\scriptsize
  \begin{tabular}{|l|r|r|}\hline 
{\bf Survey} & {\bf Papers} & {\bf Citations} \\ \hline
Sloan Digital Sky Survey (SDSS) & 3803 & 163288 \\
Hubble Deep Field (HDF) & 594 & 49360 \\
Two Micron All Sky Survey (2MASS) & 1076 & 32506 \\
Great Observatories Origins Deep Survey (GOODS) & 567 & 30331 \\
Two degree Field Galaxy Redshift Survey (2dFGRS) & 206 & 25301 \\
Centre for Astrophysics (CfA) & 342 & 19883 \\
Galaxy Evolution Explorer (GALEX) & 563 & 15532 \\
Cosmic Evolution Survey (COSMOS) & 394 & 13060 \\
Ultra Deep Field (UDF) & 202 & 9169 \\
Canada France Redshift Survey (CFRS) & 90 & 8797 \\
Deep Evolutionary Exploratory Probe (DEEP/DEEP2) & 154 & 7767 \\
Classifying Objects by Medium-Band Observations (COMBO-17/COMBO17) & 89 & 6882 \\
UKIRT Infrared Deep Sky Survey (UKIDSS) & 135 & 4661 \\
VIMOS VLT Deep Survey (VVDS) & 83 & 4531 \\
Las Campanas Redshift Survey (LCRS) & 96 & 4197 \\
Southern Sky Redshift Survey (SSRS/SSRS2) & 89 & 4151 \\
APM Galaxy Survey (APMGS) & 57 & 4109 \\
Point Source Catalogue Survey (PSCz) & 95 & 3553 \\
2dF QSO redshift Survey (2QZ) & 57 & 3305 \\
Canadian Network for Observational Cosmology Field Galaxy Redshift Survey (CNOC2) & 65 & 3169 \\
CFHT Legacy Survey (CFHTLS) & 87 & 3151 \\
Galaxy Evolution from Morphologies and SEDs (GEMs) & 68 & 3109 \\
NOAO Deep Wide Field Survey (NDFWS) & 76 & 2772 \\
zCOSMOS & 60 & 2218 \\
Millennium Galaxy Catalogue (MGC) & 38 & 1854 \\
The 2dF-SDSS LRG And QSO Survey (2SLAQ) & 36 & 1636 \\
AGN and Galaxy Evolution Survey (AGES) & 34 & 1575 \\
Gemini Deep Deep Survey (GDDS) & 20 & 1479 \\
Stromlo-APM Redshift Survey (SARS) & 19 & 1402 \\
Baryonic Acoustic Oscillations Survey (BOSS) & 45 & 980 \\
ESO Slice Project (ESP) & 24 & 971 \\
The 6dF Galaxy Survey (6dFGS) & 26 & 749 \\
WiggleZ & 26 & 646 \\ 
Galaxy And Mass Assembly (GAMA) & 33 & 468 \\ 
2MASS Redshift Survey (2MRS) & 19 & 413 \\
VIMOS Public Extragalactic Redshift Survey (VIPERS) & 3 & 3 \\ \hline
  \end{tabular}
  }
 \end{center}
\vspace{1mm}

\noindent Note these numbers were determined on $28^{th}$ Dec 2012
using the SOA/NASA ADS Astronomy Query Form by searching for refereed
papers which contained abstract keywords based on the following
boolean logic: (galaxy or galaxies) and (``$<$long survey name$>$'' or
$<$short survey name$>$). The table is purely indicative and not
weighted by survey age or effective cost and is updated on an annual
basis. My apologies in advance for the many surveys not included.
\end{table}

\section{The GAMA panchromatic campaign}
As stated earlier a number of major imaging surveys are underway on a
variety of internationally leading facilities which include within
their survey footprints the GAMA sky regions.  These complimentary
imaging surveys are a mixture of GAMA motivated/led programs (e.g.,
GALEX, GMRT), those in which GAMA team members are playing a leading
role (e.g., VISTA VIKING, ASKAP DINGO), to those in which GAMA is
peripheral or playing a supporting role (e.g., VST KIDs, WISE,
Herschel-Atlas). Regardless of the motivation these data collectively
form the GAMA panchromatic campaign and need to be assembled,
astrometrically matched, and for photometric measurements to be
conducted which manages highly variable flux limits and spatial
resolutions. Typically only 10-20\% of the GAMA galaxies have clear
unambiguous counterparts in all bands and spatial resolutions which
vary with wavelength from 0.5$''$ to 30$''$. The significance of
managing these two issues (flux limits and spatial resolutions),
cannot be overstated. The correct strategy to manage this process is
not yet entirely clear, the wrong strategy however is. To simply
conduct independent source finding in each band and then attempt to
table match the resulting catalogues leads to a highly biased (ill
defined) sample with high levels of blended contamination in the
poorer resolution bands. The problem is particularly acute when
matching data from the optical to the far-IR where the incidence of
high-z interlopers and lensed sources becomes a significant
problem. Fig.~\ref{fig1} highlights some of these issues by showing
the assembled data for a single object which is clearly detected and
unambiguous but also highlights the varying spatial resolution across
the wavebands sampled. The red curve on Fig.~\ref{fig1} shows the
simple best-SED fit to these data from the MAGPHYS energy balance
software of \cite[da Cunha, Charlot \& Elbaz (2008)]{daCunha08}.  The
blue curve shows the actual energy production within this galaxy. In
particular it is worth noting the significant change in the slope of
the SED in the blue optical region where the observed colour is often
erroneously used to divide the galaxy population into {\it
  intrinsically} red and blue populations. Indeed this object is
observationally red with a high-S\'ersic index yet also dusty and
star-forming (i.e., {\it intrinsically} blue). This system alone
highlights that galaxies do not simply come in two flavours but are a
medley of intermingled parameters with all possibilities existing in
colour/star-formation/profile-shape/opacity/halo/stellar mass
parameter space. In order to complete this analysis not only are broad
spectrum measurements vital but in the case of non-detections robust
upper limits with meaningful errors. This demands the definition of
the sample in a single band from which measurements/upper-limits are
garnered across the full electro-magnetic range. A key question is
which band to start from, possibly the most logical is that most
closely aligned to stellar mass, i.e., $H$ or $K$ band.

\pagebreak

\begin{figure}[h]
\includegraphics[height=5.5in,width=4.0in,angle=-90.0]{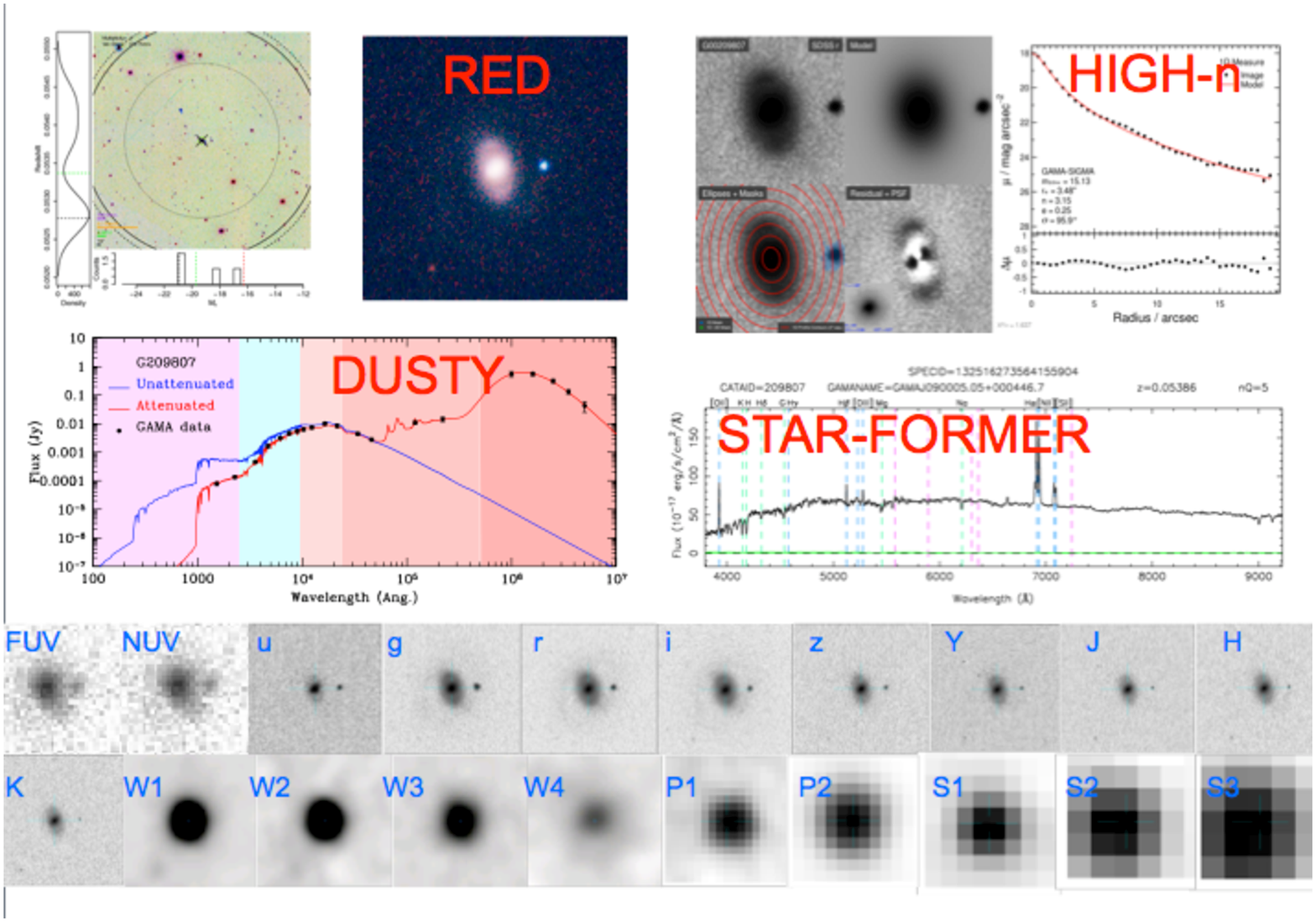}

\vspace{-0.5cm}

\caption{The information available for our 380,000
  galaxies. Which includes (clockwise from top-right): halo and group
  information, colours, 2D profiles (single S\'ersic and bulge-disc
  decompositions), spectroscopic line analysis, panchromatic imaging,
  and UV-farIR SED modelling. \label{fig2}}

\vspace{-0.75cm}

\end{figure}

\section{Summary}
Unlocking the interdependency of galaxy properties as a function of
environment and redshift requires a comprehensive panchromatic survey
such as GAMA. Fig.~\ref{fig2} shows the type of data we are now
assembling for 380,000 galaxies which will enable us to move beyond
simple two-dimensional descriptions of the galaxy population and allow
us to explore the rich interdependencies of the many properties which
we will measure. Extensions to fainter flux-limits, higher spatial
resolutions, and higher redshifts are underway.

\end{document}